\documentclass{article}
\pdfoutput=1
\usepackage{graphicx}
\usepackage{geometry}
\usepackage{array}
\usepackage{amsmath}
\begin{document}

\title{Comparative study of Wavelet transform and Fourier domain filtering for medical image denoising} 
\author{M. Ali Saif$^1$\footnote{masali73@gmail.com}, Bassam M. Mughalles$^2$\footnote{bassammughales@gmail.com} and Ibrahim G. H. Loqman$^2$\footnote{i.luqman@su.edu.ye}\\
$^1$ Department of Physics,
University of Amran,
Amran,Yemen.\\
$^2$ Department of Physics, University of Sana'a, Sana'a, Yemen.}

\maketitle
\begin{abstract}
Denoising of images is a crucial preprocessing step in medical imaging, essential for improving diagnostic clarity. While deep learning methods offer state-of-the-art performance, their computational complexity and data requirements can be prohibitive. In this study we present a comprehensive comparative analysis of two classical, computationally efficient transform-domain techniques: Discrete Wavelet Transform (DWT) and Discrete Fourier Cosine Transform (DFCT) filtering. We evaluated their efficacy in denoising medical images which corrupted by Gaussian, Uniform, Poisson, and Salt-and-Pepper noise. Contrary to the common hypothesis favoring wavelets for their multi-resolution capabilities, our results demonstrate that a block-based DFCT approach consistently and significantly outperforms a global DWT approach across all noise types and performance metrics (SNR, PSNR, IM). We attribute DFCT's superior performance to its localized processing strategy, which better preserves fine details by operating on small image blocks, effectively adapting to local statistics without introducing global artifacts. This finding underscores the importance of algorithmic selection based on processing methodology, not just transform properties, and positions DFCT as a highly effective and efficient denoising tool for practical medical imaging applications.
\end{abstract}

\section{Introduction}
The removal of additive noise from images, a process known as image denoising, remains one of the most extensively studied topics in image processing \cite{ela1,mil}. The presence of noise not only degrades visual quality but can also obscure subtle anatomical or pathological features, thereby impeding accurate diagnosis. Over the past decades, numerous of denoising techniques has been developed, broadly categorized into spatial domain, transform domain, statistical model-based approaches, and, more recently, deep neural networks.

Spatial domain methods, such as mean, median, bilateral filtering \cite{tom,ela}, and total variation \cite{bec,hu}, directly operate on pixel values intensities to smooth noise while attempting to preserve edges. In contrast, transform domain methods represent an image in an alternative basis such as Fourier \cite{roa,ahm}, or wavelet transforms \cite{xiz,dau,che} where noise and signal components can be more effectively separated. Statistical approaches, including Wiener filtering and Bayesian methods, exploit prior knowledge of noise statistics and image models to achieve adaptive noise suppression \cite{pra,ric,aga}. Although deep learning-based methods have recently achieved remarkable denoising performance (See Ref. \cite{ela1} and the refernce therin), however they often require significant computational resources, large annotated datasets, and considerable expertise to implement and train.

In this study we focus on two robust and computationally efficient transform-domain methods: Wavelet transformation and Fourier filtering. While deep learning excels in performance, the complexity and resource demands make simpler, well-established methods highly valuable for many practical scenarios. Wavelet-based denoising is particularly renowned for its multi-resolution analysis capability, enabling effective noise suppression while retaining fine structural details \cite{agr,kum}. Fourier-based filtering provides a global frequency perspective, making it exceptionally suitable for removing periodic or high-frequency noise patterns \cite{roa,ahm}.

The need for such reliable denoising is paramount in medical imaging modalities like Computed Tomography (CT) and Magnetic Resonance Imaging (MRI). CT images are often degraded by Poisson noise due to low-dose acquisition protocols aimed at minimizing patient radiation exposure. Conversely, MRI images are frequently contaminated by Gaussian or Rician noise originating from thermal fluctuations and imperfections in RF coils. In additional to that, noise can arise from various sources, including photon counting statistics, electronic sensor fluctuations, and patient motion. The presence of noise not only degrades visual quality but can also obscure subtle anatomical or pathological structures, thereby impacting diagnostic accuracy. Applying advanced yet efficient denoising algorithms in these domains is crucial for enhancing image quality, improving tissue contrast, and ultimately supporting more accurate diagnostic decisions \cite{hoc}.

In this work we aim to provide a systematic comparison of Wavelet transform and Fourier domain filtering for medical image denoising. We describe the theoretical foundations of each method, implement them on sample CT and MRI data, and evaluate their performance using standard image quality metrics.

\section{Methods describtion}
\label{sec:methods}

Medical image denoising has been an active research area for decades, driven by the imperative to improve diagnostic accuracy while reducing patient exposure to radiation or shortening scan times. Conventional low-pass and smoothing filters can suppress noise but often blur fine details, which are crucial for detecting small lesions or subtle tissue changes. Consequently, more sophisticated denoising strategies are required to optimally balance noise reduction with detail preservation.

A digital imeges can be represented as matrix $\mathbf{z}$ of dimension $N\times M$, where ${z(i): i\in I}$ is a pixel value and $I=\left\{1,...,N\right\}\times\left\{1,...,M\right\}$ is the spatial domain. The noisy observation of this image is given by:
\begin{eqnarray}
s(i)=z(i)+\eta(i)
\end{eqnarray}
Where $z(i)$ is the clean image pixel and $\eta(i)$ represents additive noise. The common types of additive noise include Gaussian, uniform, and Poisson noise. This model assumes that noise is distributed throughout the image; that is, each pixel in the noisy image is the sum of the true pixel value and an independent and identically distributed ($i.i.d.$) random variable.

\subsection{Wavelet Transform Denoising}
\label{subsec:wavelet_denoising}
Wavelet-based methods have emerged as one of the most effective approaches for image denoising due to their ability to provide simultaneous spatial and frequency localization. The foundational work of Donoho and Johnstone \cite{dono} on wavelet thresholding laid the groundwork for soft and hard thresholding techniques, where noise-dominated coefficients are suppressed while signal-dominated coefficients are retained \cite{med,tas,buc}.

The wavelet transform is a mathematical tool that represents a signal or image in terms of small wave-like functions called wavelets, which are localized in both space and frequency. Unlike the Fourier transform, which provides only global frequency content, the wavelet transform offers a multi-resolution analysis, capturing both coarse structures and fine details effectively.

The Discrete Wavelet Transform (DWT) is a common implementation for images. The 2D-DWT decomposes an image by passing it through low-pass and high-pass filters along the rows and columns, producing four sub-bands:
\begin{itemize}
    \item \textbf{LL}: Low frequencies in both directions (approximation component)
    \item \textbf{LH}: Low horizontal, high vertical frequencies (vertical edges)
    \item \textbf{HL}: High horizontal, low vertical frequencies (horizontal edges)
    \item \textbf{HH}: High frequencies in both directions (diagonal details and noise)
\end{itemize}

The LL sub-band can be recursively decomposed to achieve multiple scales of analysis, as illustrated in Fig. 1. The efficacy of DWT for denoising stems from the tendency of noise to spread across the high-frequency sub-bands (LH, HL, HH), while important image structures are concentrated in the LL sub-band and strong coefficients within the detail sub-bands. The standard denoising procedure involves transforming the image, applying a thresholding rule to the detail coefficients (shrinking or removing those likely to be noise), and then reconstructing the image from the modified coefficients. This approach generally preserves edges and textures better than simple spatial smoothing.

\begin{figure}[htp]
\centering
\includegraphics[width=50mm,height=50mm]{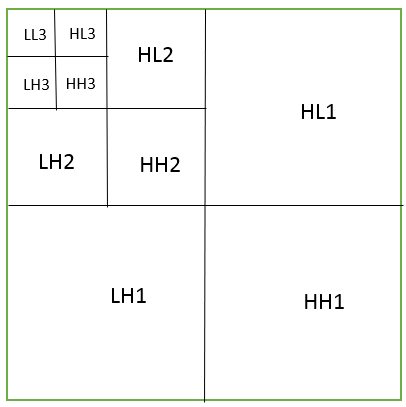}
\caption{Three-level decomposition of an image into sub-bands using a 2D Discrete Wavelet Transform.}  
\end{figure}

\subsubsection{Types of wavelet filters}
\label{subsubsec:wavelet_types}
Numerous wavelet families have been used in DWT, each with properties suited to different applications \cite{zou,coi,str}. The most prominent filters, which are employed in this study, are listed below:

\begin{itemize}
    \item \textbf{Haar Wavelet}: The simplest and fastest orthogonal wavelet, but can produce blocky artifacts due to its discontinuous nature.
    \item \textbf{Daubechies Wavelet($n$)}: A family of orthogonal wavelets characterized by order $n$ of vanishing moments with $n>2$, offering a good trade-off between smoothness and localization. Higher orders ($n$) provide increased smoothness.
    \item \textbf{Coiflet Wavelet($n$)}: Designed to be more symmetric than Daubechies wavelets, with scaling functions that also have vanishing moments of $1\leq n \leq 5$. Often used for their symmetrical properties. 
    \item \textbf{Symlet Wavelet($n$)}: A family of nearly symmetric orthogonal wavelets with $n>4$, also known as the least asymmetric wavelets, designed to improve symmetry compared to Daubechies wavelets.
    \item \textbf{CDF 9/7 Wavelet}: A biorthogonal wavelet (Cohen-Daubechies-Feauveau) known for its symmetry and use in the JPEG2000 compression standard. Excellent for image reconstruction.
    \item \textbf{Biorthogonal Spline Wavelet (BIOS($n$,$m$))}: A family of biorthogonal wavelets offering linear phase (symmetry) in which $n+m$ is even. It is beneficial for image processing tasks like denoising and reconstruction, often used in medical imaging.
    \item \textbf{Meyer Wavelet}: An orthogonal wavelet defined in the frequency domain, offering excellent frequency localization but is less common in discrete implementations.
    \item \textbf{Shannon Wavelet}: Also defined in the frequency domain, ideal in theory but suffers from slow decay in the time domain, making it less practical for discrete applications.
\end{itemize}

\subsubsection{Thresholding functions}
The core of wavelet-based denoising involves transforming the image into the wavelet domain. In this domain, noise tends to appear as small coefficients scattered everywhere, while important features (edges, texture) appear as large coefficients. Thresholding is applied to remove noise (small coefficients) while preserving the signal (large coefficients). Thresholding is performed on the detail coefficients at each level of the wavelet decomposition. Table~\ref{tab:threshold_funcs} summarizes the key thresholding functions used in this study.

\begin{table}[htb]
    \centering
    \caption{Thresholding functions used for wavelet denoising. For all functions, $\delta > 0$ is the threshold parameter. For SmoothGarrote, $n$ is a positive real number.}
    \label{tab:threshold_funcs}
    \begin{tabular}{|l|c|}
        \hline
        \textbf{Threshold Name} & \textbf{Function} \\
        \hline
        Hard &
        $\theta_H(x) =
        \begin{cases}
            0 & \text{if } |x| \leq \delta \\
            x & \text{if } |x| > \delta
        \end{cases}$ \\
        \hline
        Soft &
        $\theta_S(x) =
        \begin{cases}
            0 & \text{if } |x| \leq \delta \\
            \text{sgn}(x)(|x| - \delta) & \text{if } |x| > \delta
        \end{cases}$ \\
        \hline
        Smooth Garrote & $\theta_{SG}(x) = \dfrac{x^{2n+1}}{x^{2n} + \delta^{2n}}$ \\
        \hline
        Piecewise Garrote &
        $\theta_{PG}(x) =
        \begin{cases}
            0 & \text{if } |x| \leq \delta \\
            x - \dfrac{\delta^2}{x} & \text{if } |x| > \delta
        \end{cases}$ \\
        \hline
        Hyperbola &
        $\theta_{Hyp}(x) =
        \begin{cases}
            0 & \text{if } |x| \leq \delta \\
            \text{sgn}(x)\sqrt{x^2 - \delta^2} & \text{if } |x| > \delta
        \end{cases}$ \\
        \hline
    \end{tabular}
\end{table}

Choosing the appropriate threshold value $\delta$ is crucial. A threshold that is too high leads to oversmoothing and loss of image detail, while a threshold that is too low results in insufficient noise removal. Below, we explore several prominent algorithms for calculating the optimal threshold parameter $\delta$. 

\begin{enumerate}
    \item \textbf{Universal Threshold}: This is a statistical rule for choosing the cutoff \cite{dono, ruik}. It is designed to minimize the risk of retaining noise while preserving signal details. The universal threshold is defined as:
    \begin{equation}
    \delta = \hat{\sigma} \sqrt{2 \log(N)}
    \end{equation}
    where $N$ is the total number of pixels (or coefficients at a given sub-band), and $\hat{\sigma}$ is the estimated noise standard deviation. The noise variance $\hat{\sigma}$ is typically estimated using the median absolute deviation (MAD) of the finest-scale detail coefficients:
    \begin{equation}
    \hat{\sigma} = \frac{\text{median}(|g_i|)}{0.6745}
    \end{equation}
    where $g_i$ corresponds to the coefficients in the HH sub-band of the first decomposition level. The universal threshold provides a good starting point and is asymptotically optimal for minimizing the maximum risk \cite{dono,chan}.

    \item \textbf{SURE Threshold}: Stein's Unbiased Risk Estimate (SURE) \cite{dono,van,ste} provides a threshold that minimizes the mean squared error (MSE) between the denoised signal and the true signal. Unlike the universal threshold, SURE adapts to the actual observed wavelet coefficients, often resulting in smaller thresholds and better detail preservation. The SURE criterion for a threshold $t$ is given by:
\begin{equation}
\text{SURE}(t) = NM\sigma^2 + \sum_{i=1}^{NM} \left[ \min(|y_i|, t)^2 - 2\sigma^2 \cdot \textit{I}(|y_i| \leq t) \right]
\end{equation}
    where $y_i$ are the wavelet coefficients, $\textit{I}$ is the indicator function, and $\sigma^2$ is the noise variance. The SURE threshold $\delta_{\text{SURE}}$ is the value of $t$ that minimizes $\text{SURE}(t)$. This method is particularly useful when the noise variance is known or can be accurately estimated.

    \item \textbf{GCV Threshold}: The Generalized Cross-Validation (GCV) threshold is similar to SURE but does not require prior knowledge of the noise variance. It automatically selects an appropriate threshold by minimizing a cross-validation score that balances bias and variance. The GCV criterion for a threshold $t$ is defined as:
    \begin{equation}
    \text{GCV}(t) = \frac{ \frac{1}{NM} \sum_{i=1}^{NM} (y_i - \hat{y}_i(t))^2 }{ \left(1 - \frac{N_0(t)}{NM}\right)^2 }
    \end{equation}
    where $\hat{y}_i(t)$ are the thresholded coefficients, and $N_0(t)$ is the number of coefficients set to zero by threshold $t$. The optimal threshold $\delta_{\text{GCV}}$ minimizes $\text{GCV}(t)$.

    \item \textbf{Level-Dependent Thresholds}: Both the Universal and SURE principles can be applied in a level-dependent manner \cite{ari}. A Universal-Level threshold calculates $\delta_j$ separately for each decomposition level $j$ using Eq. (2) with $N$ being the number of coefficients at that level. Similarly, a SURE-Level threshold (often called SureShrink) finds the optimal $\delta_j$ for each level by minimizing the SURE criterion for the coefficients at that level. Level-dependent thresholds can reduce oversmoothing by applying less aggressive thresholding at coarse scales (preserving structures) and more aggressive thresholding at fine scales (suppressing noise).
\end{enumerate}
  
\subsection{Fourier Domain Filtering}
\label{subsec:fourier_denoising}

Fourier transform-based denoising exploits the fact that noise and signal components often occupy distinct frequency bands. Traditional low-pass Fourier filters are effective in removing high-frequency noise, while band-stop filters can target periodic noise patterns. However, standard Fourier methods lack spatial localization, which may lead to the loss of localized features and the creation of global artifacts like ringing. Hybrid approaches that combine Fourier filtering with wavelet decomposition have been proposed to overcome this limitation \cite{roa}. In this work, we employ a block-based Discrete Fourier Cosine Transform (DFCT) approach, which mitigates these issues by operating locally.

Fourier filtering works by transforming the image from the spatial domain (pixels) to the frequency domain (sinusoidal components). In the frequency domain:
\begin{itemize}
    \item Low frequencies correspond to large, smooth image structures.
    \item High frequencies correspond to fine details and noise (especially random noise).
\end{itemize}
By selectively modifying specific frequency components, we can suppress noise while keeping important structures.

We utilize the 2D Discrete Fourier Cosine Transform (DFCT) to decompose the image into its frequency spectrum. The DFCT produces real-valued coefficients, simplifying calculation and interpretation compared to the complex-valued Discrete Fourier Transform (DFT). The DFCT of an image $s(i), i=(i_1,i_2)\in I$ is defined as \cite{ahm,kha}:
\begin{equation}
F(k_1, k_2) = \frac{1}{\sqrt{NM}} \sum_{i_1=0}^{N-1} \sum_{i_2=0}^{M-1} s(i_1, i_2) \cos\left( \frac{\pi}{N} (i_1+\frac{1}{2}) k_1 \right) \cos\left( \frac{\pi}{M} (i_2+\frac{1}{2}) k_2 \right)
\end{equation}
for $k_1 = 0, \ldots, N-1$, $k_2 = 0, \ldots, M-1$.

To design a filter in the frequency domain, a mask $H(k)$ is created to determine how much of each frequency component is retained. Common filter types include:
\begin{enumerate}
    \item \textbf{Low-pass filter}: Keeps low frequencies by attenuating coefficients above a certain frequency cutoff. Effective for removing high-frequency noise but can blur edges.
    \item \textbf{High-pass filter}: Keeps high frequencies and removes low frequencies. Useful for edge enhancement but not typically for denoising.
    \item \textbf{Band-pass filter}: Targets a specific range of frequencies.
    \item \textbf{Notch filter}: Removes noise at specific frequency points (e.g., for removing periodic interference).
\end{enumerate}
In this study, we primarily employ a simple low-pass hard thresholding filter, defined by the mask:
\begin{equation}
H(k) =
\begin{cases}
1 & \text{if } |F(k)| \leq \tau \\
0 & \text{otherwise}
\end{cases}
\end{equation}
where $\tau$ is the frequency threshold. The filtering operation in the Fourier domain is a simple multiplication:
\begin{equation}
G(k) = H(k) \cdot F(k)
\end{equation}
where $G(k)$ is the filtered spectrum.

After modifying the spectrum, the image is converted back to the spatial domain using the Inverse Discrete Fourier Cosine Transform (IDFCT).

To avoid artifacts associated with global Fourier processing and to better adapt to local image content, the DFCT is applied in a block-based manner. The image is divided into small, overlapping $16 \times 16$ blocks \cite{wol1}. The DFCT denoising procedure is applied independently to each block, and the final denoised image is formed by averaging the overlapping results from all blocks. This approach provides spatial adaptation and helps prevent the ringing artifacts typical of global Fourier filtering.

\section{Performance Evaluation}
\label{sec:performance}

To quantitatively assess the efficacy of the denoising filters and facilitate a comparison between them, we adopt the following standard image quality metrics.

\begin{enumerate}
    \item \textbf{Signal-to-Noise Ratio (SNR):} This metric measures the ratio of the power of the original signal to the power of the noise. A higher SNR indicates a stronger signal relative to the noise and thus better denoising performance. For a clean image $s$ and its denoised version $\hat{s}$, the SNR is calculated as:
    \begin{equation}
    \text{SNR} = 10 \log_{10} \left( \frac{\text{Var}(s)}{\text{Var}(\hat{s} - s)} \right)
    \end{equation}
    where $\text{Var}(\cdot)$ denotes the variance of the image.

    \item \textbf{Peak Signal-to-Noise Ratio (PSNR):} This is a widely used metric, especially in image processing, that defines the ratio between the maximum possible power of a signal and the power of corrupting noise. A higher PSNR value indicates better denoising quality. It is defined as:
    \begin{equation}
    \text{PSNR} = 10 \log_{10} \left( \frac{(\max(|s|))^2}{\text{MSE}} \right)
    \end{equation}
    where $\text{MSE}$ is the Mean Squared Error between the clean image $s$ and the denoised image $\hat{s}$, given by $\text{MSE} = \frac{1}{NM}\sum_{i \in I} (s(i) - \hat{s}(i))^2$, and $\max(|s|)$ is the maximum possible pixel value of the image (e.g., 255 for an 8-bit image).

    \item \textbf{Image Distance (IM):} This metric calculates the Euclidean distance between the clean and denoised images in the vector space. It provides a direct measure of the overall error magnitude. A lower IM value signifies that the denoised image is closer to the original and therefore denotes better performance. It is defined as:
    \begin{equation}
    \text{IM} = \| s - \hat{s} \|_2 = \sqrt{ \sum_{i \in I} (s(i) - \hat{s}(i))^2 }
    \end{equation}
\end{enumerate}

\section{Denoising Results}
\label{sec:results}

\begin{figure}[htp]
\centering
\mbox{[a)Clean]{\includegraphics[height=30mm]{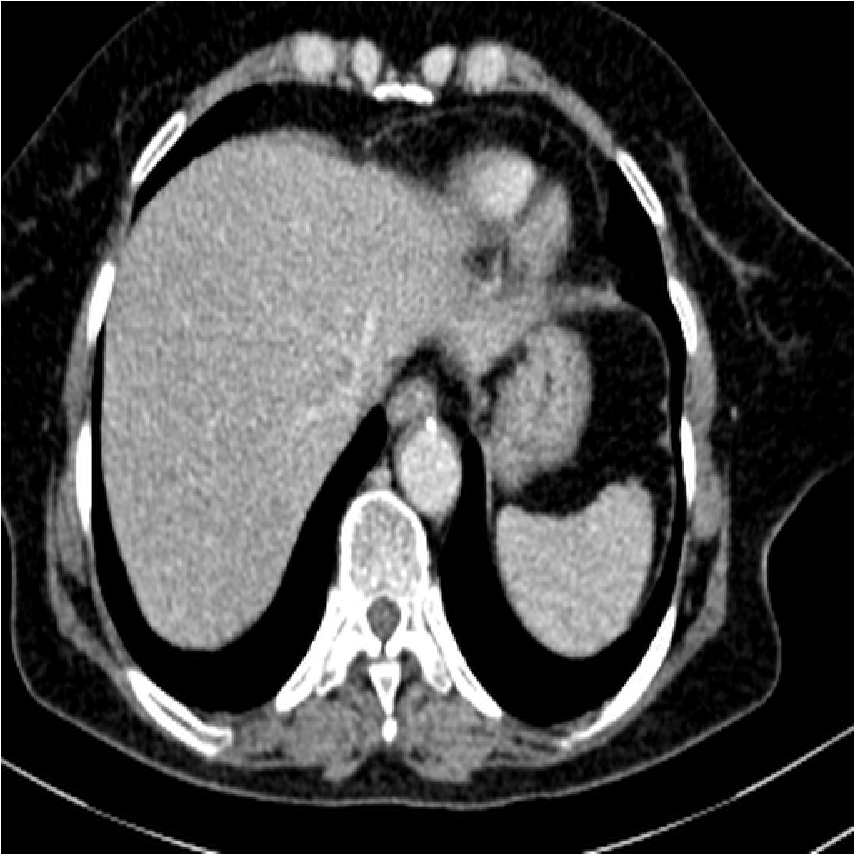}}\quad
[b)Gaussian]{\includegraphics[height=30mm]{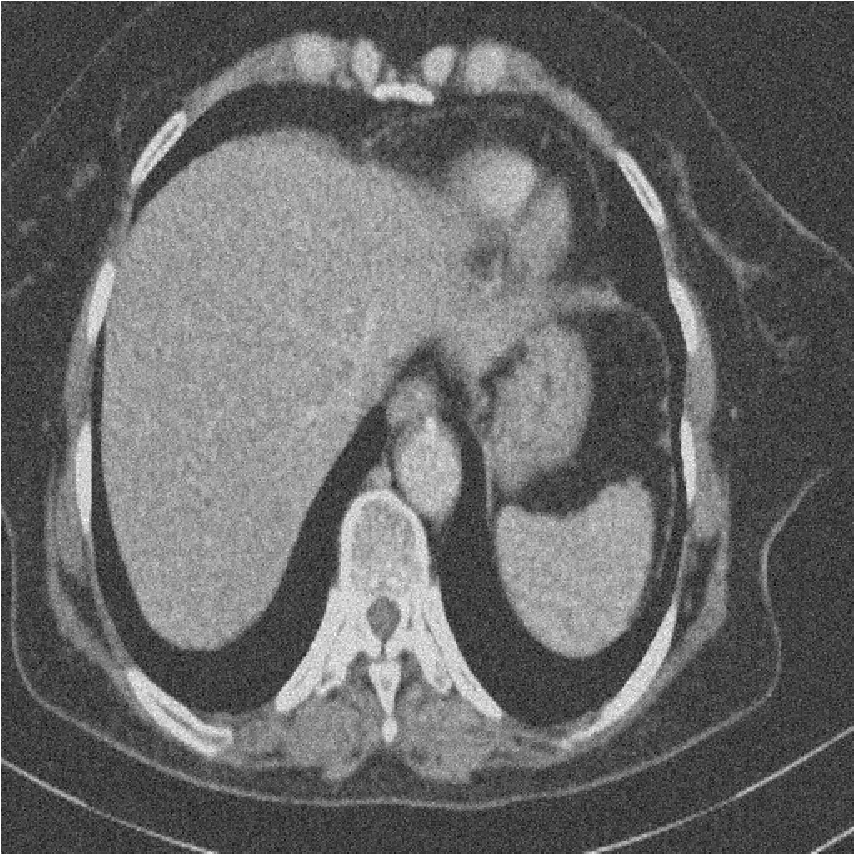}}\quad
[c)Uniform]{\includegraphics[height=30mm]{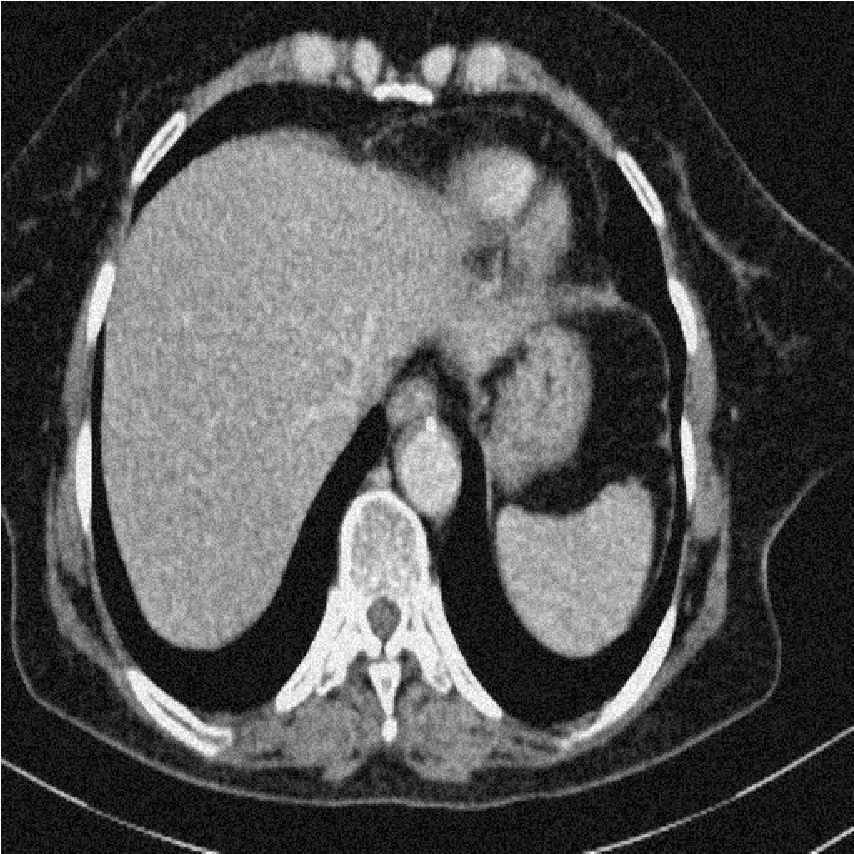}}}
\mbox{[d)Poisson]{\includegraphics[height=30mm]{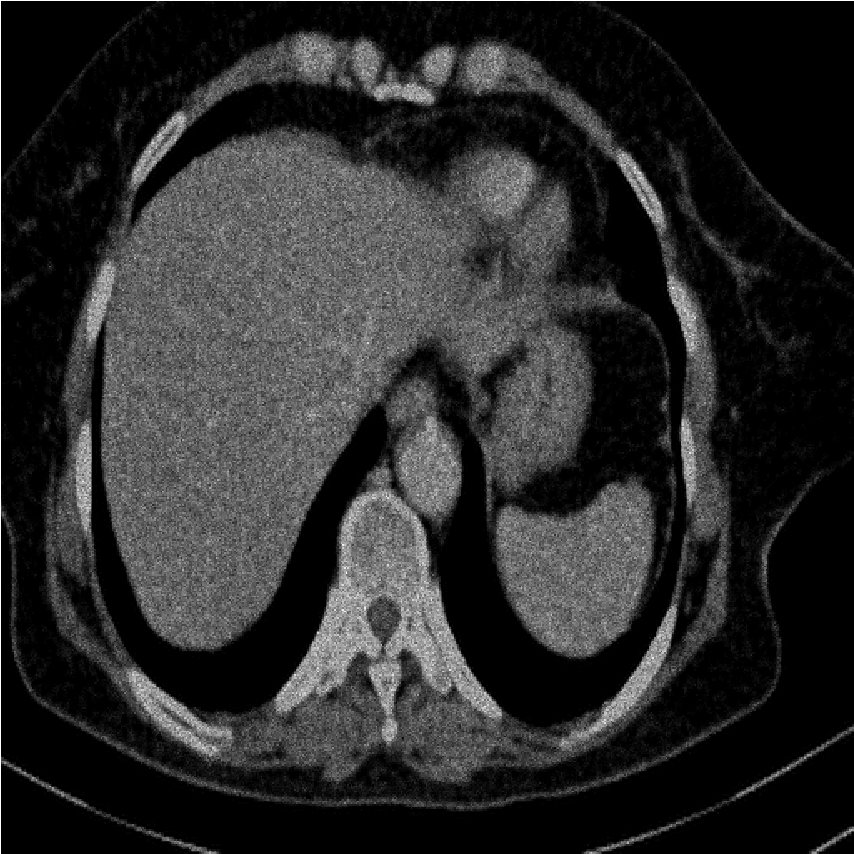}}\quad
[e)Salt Pepper]{\includegraphics[height=30mm]{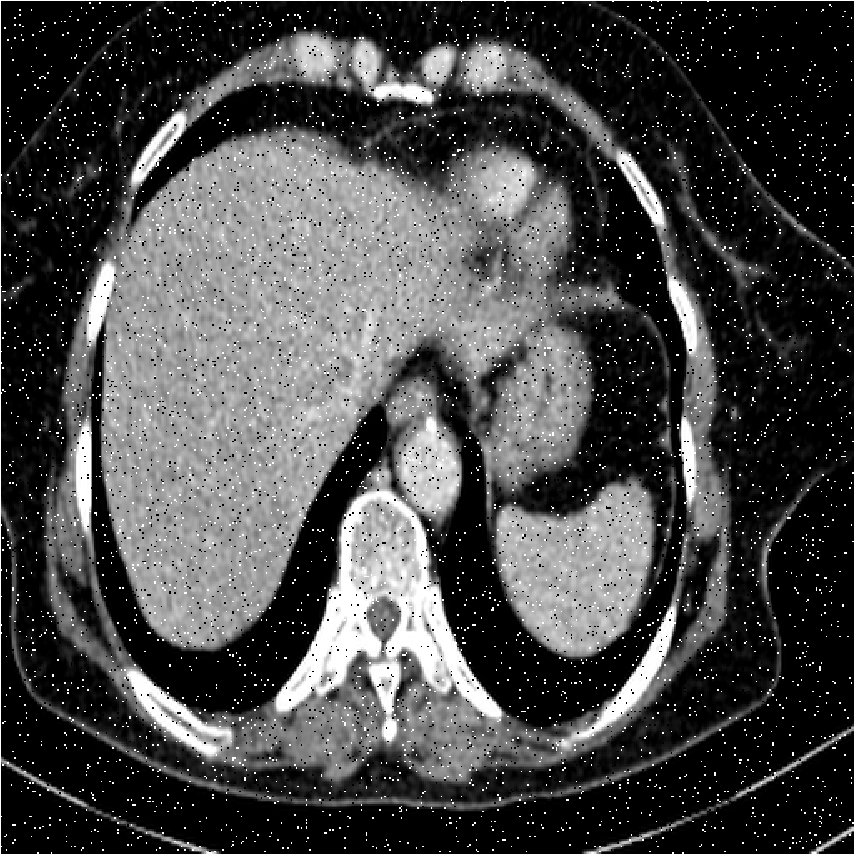}}}
\caption{Test image used for denoising evaluation: (a) original clean CT scan; and its noisy versions corrupted with (b) Gaussian noise ($\sigma=25$); (c) uniform noise; (d) Poisson noise ($\sigma=42.85$); (e) salt-and-pepper noise (density $=0.02$).}
 \label{fig:clean_img} 
\end{figure}

We performed extensive numerical simulations to examine the performance of the aforementioned denoising methods. For this purpose, we used a $512 \times 512$ pixel grayscale CT scan image, shown in Fig.~\ref{fig:clean_img}a. Noisy versions of this image were created by adding different types of noise at various strengths. The noise types considered are \textit{Gaussian noise}, \textit{uniform noise}, \textit{salt and pepper noise}, and \textit{Poisson noise}. The following subsections present and explore the results obtained from the various denoising methods. All computational programming were carried out using \textit{Mathematica 12.0}.

\subsection{Denoising using Wavelet transformation}
\label{subsec:wavelet_results}

We denoised the images using all wavelet filters mentioned in Section~\ref{subsubsec:wavelet_types} and the thresholding functions listed in Table~\ref{tab:threshold_funcs}. The specific combinations of thresholding functions and threshold selection methods used in this work are summarized in Table 2.

\begin{table}[htb]
\centering
\caption{Thresholding methods and functions evaluated in this study.}
\begin{tabular}{|l|c|c|}
\hline
\textbf{Method} & \textbf{Function} & \textbf{Threshold Method} \\
\hline
Universal & Hard & Universal \\
\hline
UniversalLevel & Hard & Universal-Level \\
\hline
VisuShrink & Soft & Universal \\
\hline
VisuShrinkLevel & Soft & Universal-Level \\
\hline
SURE & Hard & SURE \\
\hline
SureLevel & Hard & SURE-Level \\
\hline
SureShrink & Soft & SURE \\
\hline
GCV & Soft & GCV \\
\hline
GCV-Level & Soft & GCV-Level \\
\hline
SmoothGarrote & SmoothGarrote & Universal \\
\hline
PiecewiseGarrote & PiecewiseGarrote & Universal \\
\hline
Hyperbola & Hyperbola & Universal \\
\hline
\end{tabular}
\end{table}

\subsubsection{Gaussian Noise}
\label{subsubsec:gaussian_results}

Figure 2b shows the noisy image corrupted with additive white Gaussian noise at a standard deviation of $\sigma=25$ and zero mean. The initial quality metrics of this noisy image, presented in Table 3, serve as a baseline for evaluating denoising performance.

The denoising results for the image in Fig. 2b, applying various wavelet filters and threshold methods (see Table 2), are comprehensively detailed in Table 4. A clear performance hierarchy emerges from the metrics. The Meyer and Shannon wavelets demonstrated the poorest performance, failing to improve upon the noisy baseline across all thresholding functions. The Haar wavelet provided only a marginal improvement, particularly with the GCV threshold, but its overall performance remained poor compared to other filters. 

\begin{table}[htb]
\centering
\caption{Metrics values for Gaussian noisy image (Fig. 2b) relative to the clean reference (Fig. 2a).}
\begin{tabular}{|c|c|c|}
\hline
{\bf SNR}&{\bf PSNR}&{\bf IM}\\\hline
11.92&20.17&50.19\\\hline
\end{tabular}
\end{table}

In contrast, the CDF, Daubechies, Coiflet, Symlet, and Biorthogonal Spline (BIOS) wavelets exhibited strong denoising capabilities, albeit with performance dependent on the specific thresholding function. Figure 3 illustrates the PSNR values for the top four threshold functions for each of these effective wavelet filters.

The Universal threshold consistently delivered robust performance across these five wavelet families, achieving its best result (PSNR = 30.24) with the Biorthogonal Spline wavelet. The Smooth Garrote function also performed exceptionally well, yielding its highest PSNR (30.62) with the Symlet wavelet. The SURE threshold showed strong results with the CDF, Daubechies, Coiflet, and Symlet wavelets, with its peak performance (PSNR = 30.96) also occurring with the Symlet wavelet. While the GCV threshold worked well with CDF and Daubechies wavelets, its performance was inconsistent with others. The Hyperbola function proved effective with Symlet and BIOS wavelets but was less successful with other filters.

\begin{table}[htb]
\centering
\caption{Results of denoising the Gaussian noise for various wavelets filters with and thrsholding functions. The best result for each wavelet family is highlighted in bold-Part 1/2}
{\tiny
\begin{tabular}{|l|c|c|c|c|c|c|c|c|c|c|c|c|}
\hline &\multicolumn{3}{|c|}{CDF}&\multicolumn{3}{|c|}{Daubechies(12)}&\multicolumn{3}{|c|}{Coiflet(4)}&\multicolumn{3}{|c|}{Symlet(18)}\\\hline
Method&SNR&PSNR&IM&SNR&PSNR&IM&SNR&PSNR&IM&SNR&PSNR&IM\\\hline
Universal&21.24&29.49&17.15&21.35& 29.61& 16.92&21.46& 29.71& 16.72&21.89& 30.14& 15.92\\\hline
UniversalLevel&11.63&19.88&53.36&5.99& 14.25& 99.40&9.30& 17.55& 68.21&3.32& 11.57& 135.2\\\hline
VisuShrink&16.47&24.72&29.73&18.17& 26.42& 24.42&17.66& 25.92& 25.91&17.91& 26.16& 25.2\\\hline
VisuShrinkLevel&6.31&14.56&104.38&3.49& 11.75& 132.38&5.54& 13.79& 105.33&1.99& 10.25& 158.2\\\hline
SURE&{\bf 22.02}&{\bf 30.28}&{\bf 15.67}&22.19& 30.44& 15.37&22.05& 30.31& 15.62&{\bf 22.71}& {\bf 30.96}&{\bf 14.49}\\\hline
SureLevel&16.94&25.19&28.13&19.50& 27.75& 20.96&18.89& 27.15& 22.46&19.41& 27.66& 21.18\\\hline
SureShrink&17.39&25.64&26.73&18.96& 27.21& 22.30&18.31& 26.57& 24.04&18.90& 27.16& 22.47\\\hline
GCV&21.21&29.47&17.21&20.90& 29.16& 17.82&21.07& 29.33& 17.48&12.98& 21.23& 44.49\\\hline
GCV-Level&14.70&22.9&49.51&12.17& 20.43& 48.72&21.44& 29.69& 17.36&7.93& 16.18& 79.97 \\\hline
SmoothGarrote&21.29&29.54&17.05&{\bf 22.26}& {\bf 30.51}& {\bf 15.26} &{\bf 22.10}& {\bf 30.36}& {\bf 15.53}&22.36& 30.62& 15.07\\\hline
PiecewiseGarrote&19.31&27.57&21.40&20.42& 28.68& 18.84&20.24& 28.50& 19.24&20.69& 28.91& 18.34\\\hline
Hyperbola&20.33&28.58&19.05&21.02& 29.27& 17.59&20.97& 29.22& 17.69&21.38& 29.63& 16.88\\\hline
\end{tabular}}
\end{table}
\begin{table}[h]
\centering
{\tiny
\begin{tabular}{|l|c|c|c|c|c|c|c|c|c|c|c|c|}
\multicolumn{13}{l}{Denoising results for Gaussian noise-Part 2/2}\\\hline
&\multicolumn{3}{|l|}{BIOS(2,8)}&\multicolumn{3}{|c|}{Haar}&\multicolumn{3}{|c|}{Meyer(14)}&\multicolumn{3}{|c|}{Shannon}\\\hline
Method&SNR&PSNR&IM&SNR&PSNR&IM&SNR&PSNR&IM&SNR&PSNR&IM\\\hline
 \mbox{Universal}& 21.98 & 30.24 & 15.75 & 15.7 & 23.96 & 32.44 & 7.33& 15.58& 85.67 & 5.43& 13.68& 107.5\\\hline
 \mbox{UniversalLevel} & 12.03 & 20.29 & 49.58 & 6.37 & 14.63 & 94.99 & 6.73& 14.98& 96.69& 4.68& 12.93& 122.3\\\hline
 \mbox{VisuShrink} & 20.44 & 28.69 & 18.81 & 12.73 & 20.99 & 45.67 & 7.07& 15.33& 88.27& 4.95& 13.21& 113.6\\\hline
 \mbox{VisuShrinkLevel} & 7.14 & 15.39 & 87.97 & 2.92 & 11.18 & 141.3 & 4.40& 12.66& 122.3& 2.81& 11.06& 147.7\\\hline
 \mbox{SURELevel} & 15.97 & 24.23 & 31.45 & 15.97& 24.23& 31.45& 7.32& 15.57& 85.74& 5.30& 13.55& 109.1\\\hline
 \mbox{SUREShrink} & 18.2 & 26.45 & 24.35 & 9.21 & 17.46 & 68.55 & 7.28& 15.53& 86.14& 5.02& 13.28& 112.6\\\hline
 \mbox{GCV} & 20.97 & 29.23 & 17.7 & 12.95 & 21.21 & 44.52 & 7.13& 15.38& 87.75& 4.48& 12.74& 119.8\\\hline
 \mbox{GCVLevel} & 9.77 & 18.03 & 64.34 & {\bf 17.69} &{\bf 25.95} & {\bf 25.81} & 7.18& 15.44& 87.07& 4.69& 12.94& 120.8\\\hline
 \mbox{SURE} & 13.96 & 22.2 & 43.87 & 14.51 & 22.76 & 37.22 & 7.15& 15.41& 90.94& 5.44& 13.69& 107.4\\\hline
 \mbox{SmoothGarrote} & 21.99 & 30.24 & 15.75 & 16.95 & 25.2 & 28.11 & 7.29& 15.54& 86.05& 5.36& 13.61& 108.4\\\hline
 \mbox{PiecewiseGarrote} & 22.84 & 31.09 & 14.28 & 14.52 & 22.77 & 37.19 & 7.31& 15.57& 85.82& 5.34& 13.59& 108.7\\\hline
 \mbox{Hyperbola} & {\bf 23.19} &{\bf 31.44} &{\bf 13.71}& 15.209 & 23.4637 & 34.36 & 7.33& 15.59& 85.62& 5.38& 13.64& 108.1\\\hline
\end{tabular}}
\end{table}

We summarize the optimal thresholding function for each wavelet filter in Table 11. The SURE threshold achieved the best performance for the CDF and Symlet wavelets, while Smooth Garrote was optimal for Daubechies and Coiflet wavelets. Overall, the top denoising performance was achieved by the Biorthogonal Spline wavelet (BIOS(2,8)) using the Hyperbola thresholding function, attaining a PSNR of 31.44 dB.
 
\begin{figure}[htb]
\centering
\includegraphics[width=140mm,height=70mm]{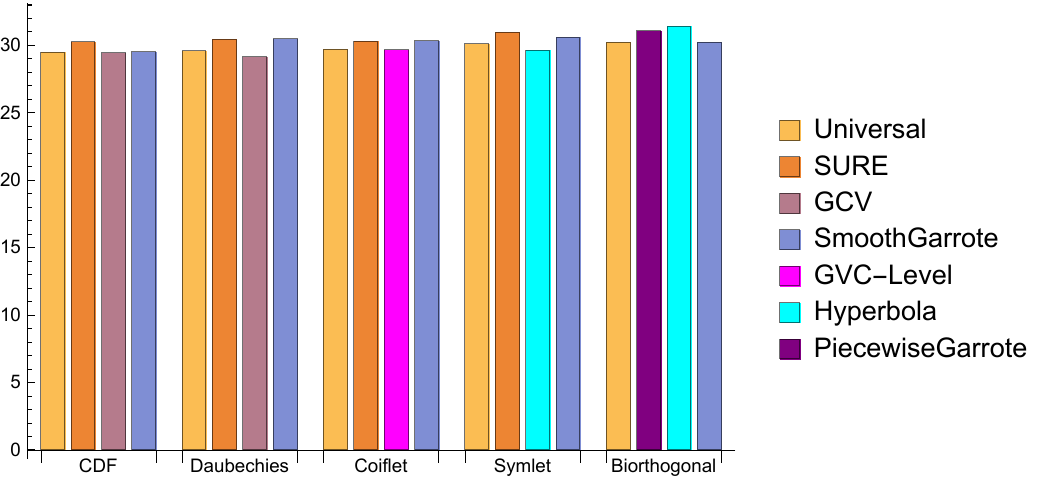}
\caption{Denoising performance (PSNR in dB) for Gaussian noise: comparison of the top four thresholding functions for each wavelet filter.} 
\end{figure}

		
\subsubsection{Uniform noise}
This section evaluates the capability of wavelet filters using different kind of thresholds to remove the uniform noise. The clean image in Fig. 2a was corrupted with uniform noise at an intensity of $\sigma=25$, resulting in the noisy image shown in Fig. 2c. The baseline metrics for this noisy image are provided in Table 5.

\begin{table}[htb]
\centering
\caption{Metrics values for uniform noisy image (Fig. 2c) relative to the clean reference (Fig. 2a).}
\begin{tabular}{|c|c|c|}
\hline
{\bf SNR}&{\bf PSNR}&{\bf IM}\\
\hline
16.68&24.93&28.99\\
\hline
\end{tabular}
\end{table}

The resuts of denoising are presented in Table 6. Mirroring the results for Gaussian noise, the Meyer and Shannon wavelets failed to denoise the image effectively. The Haar wavelet again provided only a slight improvement, with performance significantly lagging behind the other wavelets.

For the five effective wavelet filters (CDF, Daubechies, Coiflet, Symlet, BIOS), Figure 4 displays the top four performing threshold functions. The Universal, SURE, Smooth Garrote, Hyperbola, and Piecewise Garrote functions achieved the most effective denoising. As summarized in Table 11, the SURE threshold yielded the best results for the CDF, Daubechies, Coiflet, and Symlet wavelets. However, consistent with the Gaussian noise case, the Biorthogonal Spline wavelet (BIOS(2,8)) achieved the overall highest performance using the Hyperbola threshold, attaining a PSNR of 34.46 dB.

\begin{table}[htb]
\centering
\caption{Results of denoising the uniform noise for various wavelets filters with and thrsholding functions. The best result for each wavelet family is highlighted in bold-Part 1/2}
{\tiny
\begin{tabular}{|l|c|c|c|c|c|c|c|c|c|c|c|c|}
\hline
&\multicolumn{3}{|c|}{CDF}&\multicolumn{3}{|c|}{Daubechies(4)}&\multicolumn{3}{|c|}{Coiflet(3)}&\multicolumn{3}{|c|}{Symlet(8)}\\\hline
Method&SNR&PSNR&IM&SNR&PSNR&IM&SNR&PSNR&IM&SNR&PSNR&IM\\\hline
 \mbox{Universal} & 24.39 & 32.64 & 11.93 & 24.12 & 32.37 & 12.31 & 24.48 & 32.74 & 11.80 & 24.62 & 32.87& 11.62 \\\hline
 \mbox{UniversalLevel} & 12.19 & 20.44 & 50.23 & 15.32 & 23.57 & 33.91 & 11.62 & 19.87 & 52.13 & 10.08 & 18.34& 79.61\\\hline
 \mbox{VisuShrink} & 19.42& 27.67 & 21.16& 20.08& 28.34& 19.59& 20.20& 28.45& 19.33& 20.53& 28.78& 18.62\\\hline
 \mbox{VisuShrinkLevel} & 6.65& 14.90& 100.5& 9.99& 18.25& 62.61& 6.14& 14.39& 99.13& 7.22& 15.47& 95.76\\\hline
 \mbox{SURELevel} & 17.26& 25.51& 27.15& 17.26& 25.51& 27.12& 18.59& 26.84& 23.28& 19.07& 27.32 & 22.02\\\hline
 \mbox{SUREShrink} & 20.80& 29.06& 18.03& 21.97& 30.22& 15.77& 21.36& 29.61& 16.92& 22.21& 30.47 & 15.34\\\hline
 \mbox{GCV} & 10.39& 18.64& 59.88& 14.29& 22.55& 38.16& 14.79& 23.05& 36.05& 9.43& 17.69& 66.89 \\\hline
 \mbox{GCVLevel} & 14.94& 23.19& 49.15& 19.56& 27.82& 20.81& 19.50& 27.75& 22.68& 13.54& 21.80 & 56.35 \\\hline
 \mbox{SURE} &{\bf 25.56} &{\bf 33.81} & {\bf10.43} &{\bf 25.71} & {\bf 33.97} &{\bf 10.24} & {\bf 25.57} & {\bf 33.83} &{\bf 10.41} & {\bf 26.03} &{\bf 34.28} &{\bf 9.88} \\\hline
 \mbox{SmoothGarrote} & 24.74& 33.0& 11.45& 24.93& 33.18& 11.21& 25.18& 33.44 & 10.89& 25.44& 33.70& 10.57\\\hline
 \mbox{PiecewiseGarrote} & 22.51& 30.76& 14.82& 22.70& 30.96& 14.49& 22.91& 31.17& 14.14& 23.19 & 31.44& 13.70\\\hline
 \mbox{Hyperbola} & 23.52& 31.77& 13.19& 23.56& 31.81& 13.13& 23.83& 32.08& 12.73& 24.03& 32.28 & 12.44\\\hline
\end{tabular}}
\end{table}
{\tiny
\begin{flushleft}
\begin{tabular}[htb]{|l|c|c|c|c|c|c|c|c|c|c|c|c|}
\multicolumn{13}{l}{Denoising results for uniform noise-Part 2/2}\\\hline
&\multicolumn{3}{|c|}{BIOS(2,8)}&\multicolumn{3}{|c|}{Haar}&\multicolumn{3}{|c|}{Meyer(7)}&\multicolumn{3}{|c|}{Shannon}\\\hline
Method&SNR&PSNR&IM&SNR&PSNR&IM&SNR&PSNR&IM&SNR&PSNR&IM\\\hline    
 \mbox{Universal} & 25.76& 34.02& 10.18& 18.66& 26.92& 23.07& 6.95& 15.21& 89.24& 5.45& 13.70& 107.3\\\hline
 \mbox{UniversalLevel} & 12.47& 20.73& 47.12& 6.60& 14.86& 92.48& 6.54& 14.79& 97.79& 4.69& 12.94& 122.3\\\hline
 \mbox{VisuShrink} & 23.28& 31.54& 13.55& 15.21& 23.46& 34.35& 7.07& 15.32& 88.12& 5.17& 13.42 & 110.8\\\hline
 \mbox{VisuShrinkLevel} & 7.31& 15.57& 86.37& 3.05& 11.30& 139.2& 5.08& 13.33& 113.1& 2.81& 11.07 & 147.8\\\hline
 \mbox{SURELevel} & 18.97& 27.22& 22.28& 9.20& 17.45& 68.59& 6.95& 15.21& 89.21& 5.28& 13.53& 109.4 \\\hline
 \mbox{SUREShrink} & 24.64& 32.89& 11.59& 17.44& 25.69& 26.56& 7.07& 15.32& 88.12& 5.21& 13.47 & 110.3\\\hline
 \mbox{GCV} & 9.88& 18.14& 63.51& 9.76& 18.02& 64.32& 6.94& 15.20& 89.36& 5.35& 13.61& 108.5 \\\hline
 \mbox{GCVLevel} & 14.08& 22.34& 40.17& 15.14& 23.39& 34.62& 7.17& 15.42& 90.21 & 4.68& 12.93& 121.1\\\hline
 \mbox{SURE} & 24.52& 32.78& 11.75&{\bf 20.39} &{\bf 28.65} & {\bf 18.91} & 6.94& 15.19& 89.38 & 5.45& 13.70& 107.3\\\hline
 \mbox{SmoothGarrote} & 26.07& 34.33& 9.83& 19.68& 27.94& 20.52& 6.98& 15.24& 88.97& 5.42& 13.67& 107.7\\\hline
 \mbox{PiecewiseGarrote} & 25.71& 33.96& 10.25& 17.22& 25.47& 27.25& 7.01& 15.26& 88.67& 5.42 & 13.67& 107.7\\\hline
 \mbox{Hyperbola} & {\bf 26.21} & {\bf 34.46} & {\bf 9.68}& 18.04& 26.29& 24.80& 6.99& 15.25& 88.83& 5.44& 13.69& 107.5\\\hline
\end{tabular}
\end{flushleft}}

\begin{figure}[htb]
\centering
\includegraphics[width=140mm,height=70mm]{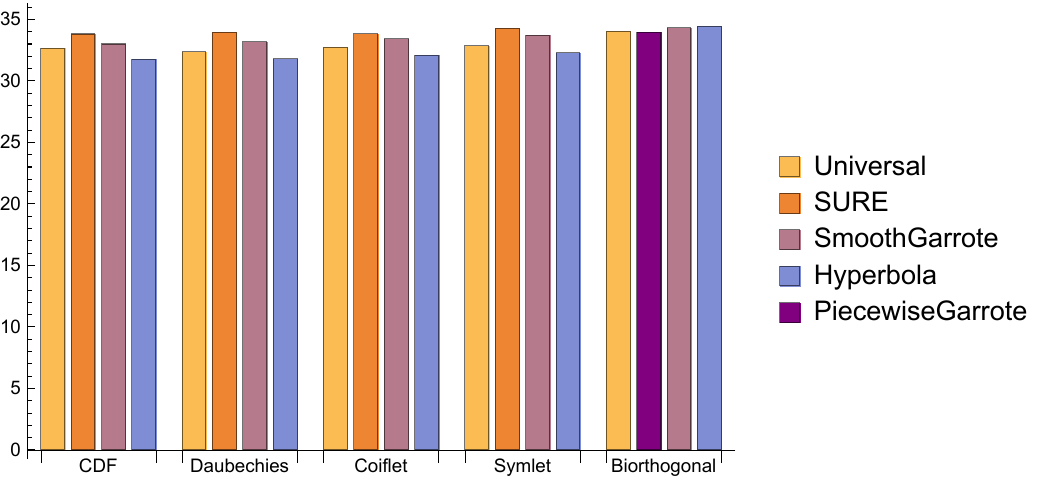}
\caption{Denoising performance (PSNR in dB) for uniform noise: comparison of the top four thresholding functions for each wavelet filter.} 
\end{figure}

%
\subsubsection{Poisson noise} 
Poisson noise, where each pixel is an independent random variable drawn from a Poisson distribution, is a common model for photon-limited imaging. A Poisson noisy image was generated from the clean reference (Fig. 2a) with a scale parameter of $\sigma=42.85$, shown in Fig. 2d. The initial quality metrics (SNR, PSNR and IM) for this image are given in Table 7.

\begin{table}[htb]
\centering
\caption{Metric values of Poissonian noisy image (Fig. 2d) relative to the clean image (Fig. 2a.)} 
\begin{tabular}{|c|c|c|}
\hline
{\bf SNR}&{\bf PSNR}&{\bf IM}\\
\hline
12.67&20.93&45.97\\
\hline
\end{tabular}
\end{table}

The denoising results for this Poisson-corrupted image are detailed in Table 8. The performance trend observed with previous noise types continued: Meyer and Shannon wavelets failed entirely, the Haar wavelet offered minimal improvement however, its performance is also bad comparing to the other wavelets. The five primary wavelets (CDF, Daubechies, Coiflet, Symlet, BIOS) performed effectively.

Figure 5 shows the best-performing threshold functions for each wavelet. The SURE, SURE-Level, VisuShrink, Piecewise Garrote, and SURE-Shrink functions delivered the most effective denoising. As summarized in Table 11, the SURE threshold provided the top performance for every one of the five effective wavelet filters. Overall, the Biorthogonal Spline wavelet (BIOS(2,4)) in combination with the SURE threshold achieved the highest denoising quality, with a PSNR of 29.14 dB.
  
\begin{table}[htb]
\centering
\caption{Results of denoising the Poisson noise for various wavelets filters with and thrsholding functions. The best result for each wavelet family is highlighted in bold-Part 1/2}
{\tiny
\begin{tabular}{|l|c|c|c|c|c|c|c|c|c|c|c|c|}
\hline
&\multicolumn{3}{|c|}{CDF}&\multicolumn{3}{|c|}{Daubechies(8)}&\multicolumn{3}{|c|}{Coiflet(4)}&\multicolumn{3}{|c|}{Symlet(11)}\\\hline
Method&SNR&PSNR&IM&SNR&PSNR&IM&SNR&PSNR&IM&SNR&PSNR&IM\\\hline
\mbox{Universal} & 13.29& 21.55& 42.81& 13.76& 22.01& 40.58& 13.45& 21.70& 42.06& 13.57& 21.82 & 41.48\\\hline
 \mbox{UniversalLevel} & 9.81& 18.06& 65.22& 8.38& 16.63& 75.44& 8.06& 16.31& 78.29& 7.47& 15.72& 84.46\\\hline
 \mbox{VisuShrink} & 18.16& 26.42& 24.44& 19.43& 27.69& 21.11& 18.71& 26.97& 22.94& 19.05& 27.30& 22.07\\\hline
 \mbox{VisuShrinkLevel} & 6.29& 14.55& 104.3& 5.21& 13.46& 108.6& 5.37& 13.62& 107.3& 4.61& 12.86& 116.8\\\hline
 \mbox{SURELevel} & 17.20& 25.46& 27.33& 18.52& 26.78& 23.44& 18.99& 27.24& 22.22& 19.03& 27.28 & 22.12\\\hline
 \mbox{SUREShrink} & 15.19& 23.44& 34.44& 16.85& 25.11& 28.41& 16.60& 24.85& 29.28& 16.69& 24.94& 28.97\\\hline
 \mbox{GCV} & 14.31& 22.56& 38.09& 14.42& 22.68& 37.60& 15.24& 23.49& 34.25& 12.60& 20.86& 46.40\\\hline
 \mbox{GCVLevel} & 10.95& 19.21& 65.91& 11.42& 19.68& 53.10& 12.62& 20.88& 46.38& 8.94& 17.20& 71.21\\\hline
 \mbox{SURE} &{\bf 19.86} & {\bf 28.11} & {\bf20.11} &{\bf 20.76} &{\bf 29.02} &{\bf 18.13} & {\bf20.57} & {\bf28.83} & {\bf18.52} & {\bf20.82} &{\bf 29.07} & {\bf18.0} \\\hline
 \mbox{SmoothGarrote} & 15.15& 23.40& 34.59& 15.93& 24.18& 31.60& 15.40& 23.65& 33.60& 15.62& 23.87& 32.75\\\hline
 \mbox{PiecewiseGarrote} & 15.89& 24.14& 31.76& 17.12& 25.37& 27.57& 16.29& 24.55& 30.30& 16.64 & 24.89& 29.13\\\hline
 \mbox{Hyperbola} & 14.82& 23.07& 35.92& 15.76& 24.01& 32.23& 15.12& 23.38& 34.68& 15.38& 23.64 & 33.67\\\hline
\end{tabular}}
\end{table}
{\tiny
\begin{flushleft}
\begin{tabular}[h]{|l|c|c|c|c|c|c|c|c|c|c|c|c|}
\multicolumn{13}{l}{Denoising results for Poisson noise-Part 2/2}\\\hline
&\multicolumn{3}{|c|}{BIOS(2,4)}&\multicolumn{3}{|c|}{Haar}&\multicolumn{3}{|c|}{Meyer(6)}&\multicolumn{3}{|c|}{Shannon}\\\hline
Method&SNR&PSNR&IM&SNR&PSNR&IM&SNR&PSNR&IM&SNR&PSNR&IM\\\hline  
 \mbox{Universal} & 13.04& 21.29& 44.10& 12.93& 21.19& 44.63& 5.81& 14.06& 101.6& 4.98& 13.23& 113.0\\\hline
 \mbox{UniversalLevel} & 9.68& 17.94 & 69.44& 5.98& 14.24& 99.34& 5.72& 13.97& 106.6& 4.37& 12.63& 126.2\\\hline
 \mbox{VisuShrink} & 17.12& 25.38& 27.54& {\bf 16.17} &{\bf 24.42} & {\bf30.75} & 6.63& 14.89& 92.54& 5.23& 13.49& 109.9\\\hline
 \mbox{VisuShrinkLevel} & 6.06& 14.32& 109.2& 2.94& 11.20& 141.01& 5.08& 13.33& 112.9& 2.78& 11.04& 148.2\\\hline
 \mbox{SURELevel} & 17.47& 25.73& 26.46& 9.20& 17.45& 68.59& 6.76& 15.02& 91.17& 5.29& 13.55& 109.2\\\hline
 \mbox{SUREShrink} & 18.55& 26.80& 23.39& 11.29& 19.55& 53.89& 6.89& 15.14& 89.98& 4.85& 13.10 & 115.0\\\hline
 \mbox{GCV} & 12.77& 21.02& 45.49& 13.82& 22.08& 40.29& 5.82& 14.08& 101.4& 4.92& 13.18& 113.6 \\\hline
 \mbox{GCVLevel} & 11.24& 19.50& 65.92& 11.37& 19.63& 53.40& 5.85& 14.11& 103.7& 4.20& 12.45& 127.1\\\hline
 \mbox{SURE} &{\bf 20.88} &{\bf 29.14} &{\bf 17.88} & 14.18& 22.43& 38.66& 6.77& 15.02& 91.10& 5.41& 13.66& 107.8\\\hline
 \mbox{SmoothGarrote} & 14.46& 22.72& 37.42& 14.21& 22.47& 38.51& 6.15& 14.41& 97.73& 5.16& 13.41& 110.8\\\hline
 \mbox{PiecewiseGarrote} & 14.90& 23.15& 35.59& 14.51& 22.76& 37.22& 6.33& 14.58& 95.83& 5.25& 13.50& 109.7\\\hline
 \mbox{Hyperbola} & 14.13& 22.38& 38.89& 13.86& 22.12& 40.09& 6.15& 14.40& 97.80& 5.16& 13.42& 110.7\\\hline
\end{tabular}
\end{flushleft}}

\begin{figure}[htb]
\centering
\includegraphics[width=140mm,height=70mm]{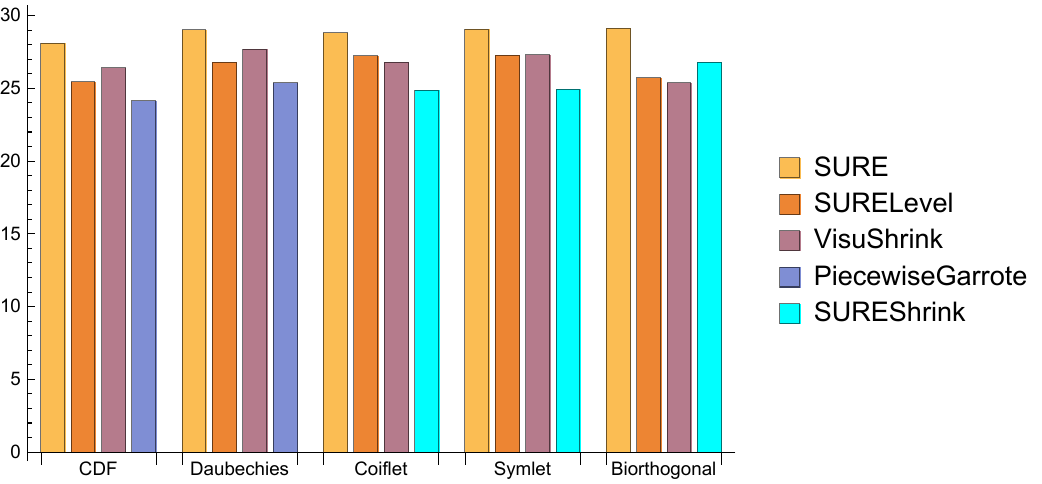}
\caption{Denoising performance (PSNR in dB) for Poisson noise: comparison of the top four thresholding functions for each wavelet filter.} 
\end{figure}

\subsubsection{Salt-and-Pepper noise}
Salt-and-pepper noise, an impulse-type noise often caused by the errors in transmission of data, manifests as sparsely occurring maximum-value (salt) and minimum-value (pepper) pixels \cite{ruik}. The clean image was corrupted with this noise using a probability of $0.02$ in both the horizontal and vertical directions, resulting in the image shown in Fig. 2e. Its baseline metrics are provided in Table 9.

The denoising results are presented in Table 10. Consistent with all previous experiments, the Meyer and Shannon wavelets failed to denoise the image, and the Haar wavelet's performance was negligible. The five main wavelet filters performed effectively, with the optimal threshold function varying per filter.

Figure 6 illustrates the top four thresholds for each of the five effective wavelets. The SURE, SURE-Level, and VisuShrink thresholds achieved good results across all wavelets. The SURE-Shrink function performed well with CDF, Symlet, and BIOS wavelets, while Smooth Garrote was effective with Daubechies and GCV with Coiflet.

Table 11 summarizes the best threshold function for each wavelet. The SURE-Level threshold achieved the best results for four of the five wavelets, while the SURE threshold was optimal for the BIOS(2,8) wavelet. The overall highest performance was attained by the Daubechies(12) wavelet using the SURE-Level threshold, achieving a PSNR of 27.0 dB.

\begin{table}[htb]
\centering
\caption{Metrics values of salt pepper noisy image (Fig. 2e) relative to the clean image (Fig. 2a.)}
\begin{tabular}{|c|c|c|}
\hline
{\bf SNR}&{\bf PSNR}&{\bf IM}\\
\hline
9.46&17.71&66.68\\
\hline
\end{tabular}
\end{table}

\begin{table}[htb]
\centering
\caption{Results of denoising the salt-and-pepper noise for various wavelets filters with and thrsholding functions. The best result for each wavelet family is highlighted in bold-Part 1/2}
{\tiny
\begin{tabular}{|l|c|c|c|c|c|c|c|c|c|c|c|c|}
\hline
&\multicolumn{3}{|c|}{CDF}&\multicolumn{3}{|c|}{Daubechies(12)}&\multicolumn{3}{|c|}{Coiflet(5)}&\multicolumn{3}{|c|}{Symlet(8)}\\\hline
Method&SNR&PSNR&IM&SNR&PSNR&IM&SNR&PSNR&IM&SNR&PSNR&IM\\\hline
 \mbox{Universal} & 9.53& 17.79& 66.10& 12.55& 20.80& 46.76& 10.02& 18.28& 62.46& 9.83& 18.08& 63.89\\\hline
 \mbox{UniversalLevel} & 7.66& 15.91& 85.81& 5.02& 13.27& 111.0& 6.41& 14.66& 94.58& 7.30& 15.55& 96.71\\\hline
 \mbox{VisuShrink} & 12.04& 20.30& 49.52& 17.55& 25.80& 26.39& 14.32& 22.57& 38.15& 13.54& 21.79& 41.71\\\hline
 \mbox{VisuShrinkLevel} & 5.13& 13.39& 115.8& 3.27& 11.52& 135.7& 4.21& 12.47& 122.0& 6.09& 14.35& 104.9\\\hline
 \mbox{SURELevel} & {\bf 16.49} & {\bf24.75} & {\bf29.75} & {\bf18.74} & {\bf27.0} &{\bf 23.07} & {\bf18.57} &{\bf 26.82} & {\bf23.50} & {\bf18.17} & {\bf26.42} &{\bf 24.60} \\\hline
 \mbox{SUREShrink} & 11.95& 20.20& 50.0& 13.51& 21.77& 41.81& 12.99& 21.24& 44.37& 12.53& 20.78 & 46.76\\\hline
 \mbox{GCV} & 10.14& 18.40& 61.5& 10.0& 18.26& 62.61& 13.87& 22.13& 40.09& 10.72& 18.98& 57.63 \\\hline
 \mbox{GCVLevel} & 8.77& 17.02& 77.49& 8.35& 16.61& 75.65& 9.90& 18.15& 63.31& 8.23& 16.49& 84.41\\\hline
 \mbox{SURE} & 16.35& 24.61& 30.24& 17.95& 26.21& 25.22& 17.52& 25.77& 26.48& 17.47& 25.72& 26.62\\\hline
 \mbox{SmoothGarrote} & 10.30& 18.55& 60.52& 14.78& 23.04& 36.18& 11.63& 19.88& 51.94& 11.11& 19.37& 55.11\\\hline
 \mbox{PiecewiseGarrote} & 10.45& 18.70& 59.48& 17.10& 25.35& 27.78& 12.25& 20.51& 48.34& 11.54 & 19.79& 52.50\\\hline
 \mbox{Hyperbola} & 10.06& 18.31& 62.23& 15.47& 23.72& 33.46& 11.35& 19.60& 53.65& 10.82& 19.08 & 56.99\\\hline
\end{tabular}}
\end{table}
{\tiny
\begin{flushleft}
\begin{tabular}[htb]{|l|c|c|c|c|c|c|c|c|c|c|c|c|}
\multicolumn{13}{l}{Denoising results for salt-pepper noise-Part 2/2}\\\hline
&\multicolumn{3}{|c|}{BIOS(2,8)}&\multicolumn{3}{|c|}{Haar}&\multicolumn{3}{|c|}{Meyer(7)}&\multicolumn{3}{|c|}{Shannon}\\\hline
Method&SNR&PSNR&IM&SNR&PSNR&IM&SNR&PSNR&IM&SNR&PSNR&IM\\\hline  
 \mbox{Universal} & 9.40& 17.65& 67.11& 9.40& 17.65& 67.11& 5.33& 13.59& 107.1& 4.56& 12.82& 117.9\\\hline
 \mbox{UniversalLevel} & 7.86& 16.12& 80.03& 5.0& 13.26& 111.2& 5.37& 13.63& 109.6& 3.99& 12.25 & 130.1\\\hline
 \mbox{VisuShrink} & 9.50& 17.75& 66.36& 9.40& 17.65& 67.11& 6.61& 14.86& 92.60& 4.85& 13.10& 114.3\\\hline
 \mbox{VisuShrinkLevel} & 5.58& 13.83& 104.5& 2.21& 10.47& 153.4& 4.79& 13.05& 115.9& 2.62& 10.87& 149.9\\\hline
 \mbox{SURELevel} & 16.96& 25.21& 28.25& 9.10& 17.35& 69.47& 6.84& 15.09& 90.18& 5.02& 13.28& 112.0\\\hline
 \mbox{SUREShrink} & 15.26& 23.52& 34.18& 9.05& 17.30& 69.86& 6.59& 14.85& 92.81& 4.23& 12.49& 122.7\\\hline
 \mbox{GCV} & 9.39& 17.64& 67.12& 9.78& 18.04& 64.22& 5.01& 13.26& 111.2& 4.26& 12.51& 122.4 \\\hline
 \mbox{GCVLevel} & 8.63& 16.88& 73.49& 9.03& 17.29& 70.0& 5.34& 13.59& 108.9& 3.72& 11.97& 132.7\\\hline
 \mbox{SURE} & {\bf18.06} & {\bf26.31} & {\bf24.95} & {\bf12.49} &{\bf 20.75} & {\bf47.04} & 6.85& 15.10& 90.09& 5.08& 13.33& 111.3\\\hline
 \mbox{SmoothGarrote} & 9.40& 17.65& 67.10& 9.40& 17.65& 67.11& 5.88& 14.13& 100.6& 4.80& 13.06 & 114.8\\\hline
 \mbox{PiecewiseGarrote} & 9.40& 17.65& 67.10& 9.40& 17.65& 67.11& 6.20& 14.45& 97.04& 4.96& 13.21& 112.8\\\hline
 \mbox{Hyperbola} & 9.40& 17.65& 67.11& 9.40& 17.65& 67.11& 5.90& 14.16& 100.3& 4.85& 13.10& 114.2\\\hline
\end{tabular}
\end{flushleft}}

\begin{figure}[htb]
\centering
\includegraphics[width=140mm,height=70mm]{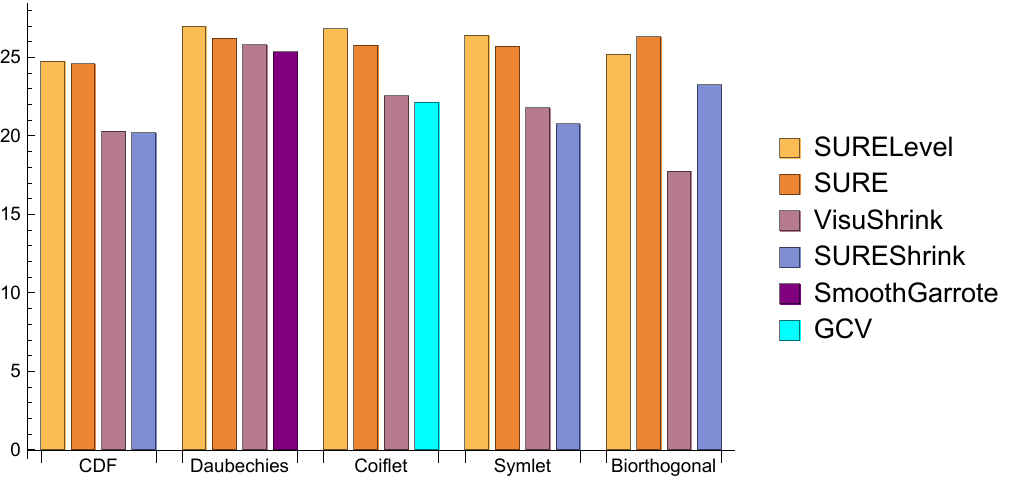}
\caption{Denoising performance (PSNR in dB) for salt-and-pepper noise: comparison of the top four thresholding functions for each wavelet filter.} 
\end{figure}

\begin{table}[htb]
\centering
\caption{Summary of optimal denoising configurations: For each noise type and wavelet family, the thresholding method achieving the highest PSNR is shown, along with the corresponding performance value.}
\begin{tabular}{|l|l|c|c|c|c|c|}
\hline
Gaussian&{\bf Wavelet}&{\bf CDF}&{\bf Daubechies(12)}&{\bf Coiflet(4)}&{\bf Symlet(18)}&{\bf BIOS(2,8)}\\\cline{2-7}
Noise&Method&SURE&SmoothGarrote&SmoothGarrote&SURE&Hyperbola\\\cline{2-7}
&PSNR&30.28&30.51&30.36&30.96&{\bf31.44}\\\hline
Uniform&{\bf Wavelet}&{\bf CDF}&{\bf Daubechies(4)}&{\bf Coiflet(3)}&{\bf Symlet(8)}&{\bf BIOS(2,8)}\\\cline{2-7}
Noise&Method&SURE&SURE&SURE&SURE&Hyperbola\\\cline{2-7}
&PSNR&33.81&33.97&33.83&34.28&{\bf34.46}\\\hline
Poisson&{\bf Wavelet}&{\bf CDF}&{\bf Daubechies(8)}&{\bf Coiflet(4)}&{\bf Symlet(11)}&{\bf BIOS(2,4)}\\\cline{2-7}
Noise&Method&SURE&SURE&SURE&SURE&SURE\\\cline{2-7}
&PSNR&28.11&29.02&28.83&29.07&{\bf 29.14}\\\hline
Salt and&{\bf Wavelet}&{\bf CDF}&{\bf Daubechies(12)}&{\bf Coiflet(5)}&{\bf Symlet(8)}&{\bf BIOS(2,8)}\\\cline{2-7}
Pepper&Method&SURELevel&SURELevel&SURELevel&SURELevel&SURE\\\cline{2-7}
&PSNR&24.75&{\bf27.00}&26.82&26.42&26.31\\\hline
\end{tabular}
\end{table}

Finally we mension to that, as it is clear from Table 11, the SURE threshold fulfills best results in $12$ case of $20$ case. However, SURELevel gives best results in four cases and SmoothGarrote and Hyperbola in two cases for each. The table shows also that, the {\bf BIOSWavelet} satisfies the highest performance in case of Gaussian noise, uniform noise and Poisson noise, however {\bf DaubechiesWavelet} gives the highest result in case of salt and pepper noise.

\subsection{Denoising using Fourier Domain Filtering}
\label{subsec:fourier_results}

Using the same noisy images from the wavelet denoising experiments (Fig. 2b-e), we present the results of denoising via the Block-Based Discrete Fourier Cosine Transform (DFCT) method described in Section~\ref{subsec:fourier_denoising}. A hard thresholding filter with a tuned threshold value $\tau$ was used for each noise type. The optimal value of $\tau$ that yielded the best denoising performance for each noise type is presented in Table 12, along with the resulting metrics.

As a reminder, the DFCT was applied locally on overlapping $16 \times 16$ blocks of the noisy image. The value of each pixel in the final denoised image is the average of the corresponding pixel in all overlapping blocks that contain it \cite{wol1}. This block-based approach is crucial for avoiding artifacts and adapting to local image statistics.

\begin{table}[htb]
\centering
\caption{Denoising performance of DFCT: Quantitative metrics before and after denoising for all noise types, using optimized threshold values $\tau$.}
\begin{tabular}{|l|l|c|c|c|}
\hline
&  & \textbf{SNR} & \textbf{PSNR} & \textbf{IM} \\
\hline
Gaussian Noise& Noisy image & 11.91 & 20.16 & 50.22 \\
& Denoised image ($\tau=0.27$) & 27.81 & 36.07 & 8.05 \\
\hline
Uniform Noise& Noisy image & 16.68 & 24.94 & 28.98 \\
& Denoised image ($\tau=0.15$) & 30.78 & 39.03 & 5.72 \\
\hline
Poisson Noise& Noisy image & 12.70 & 20.96 & 45.85 \\
& Denoised image ($\tau=0.3$) & 27.13 & 35.39 & 8.71 \\
\hline
Salt and Pepper Noise& Noisy image & 9.32 & 17.57 & 67.76 \\
& Denoised image ($\tau=0.4$) & 23.81 & 32.07 & 12.83 \\
\hline
\end{tabular}
\end{table}

\section{Discussion of results}
The Discrete Wavelet Transform (DWT) and Discrete Fourier Cosine Transform (DFCT) are two fundamental methods for image denoising. They possess intrinsic differences in how they analyze image content, which theoretically should lead to different strengths. The DWT offers good localization in both frequency and space, meaning a DWT coefficient conveys information about both the frequency content and its spatial location. In contrast, the DFCT provides excellent frequency localization but poor spatial localization within a global context; a DFCT coefficient indicates what frequency is present in a block but not its precise location within that block.

Contrary to the initial hypothesis that DWT's multi-resolution and spatio-frequency localization would grant it superior denoising capabilities \cite{xiz,che}, our results unequivocally demonstrate that the block-based DFCT approach achieved significantly higher SNR, PSNR, and lower IM values for all tested noise types (Gaussian, Uniform, Poisson, Salt-and-Pepper). Fig. 7 shows the values of PSNR for both motheds for comprision.

We posit that this superior performance is not due to a fundamental weakness of the wavelet transform itself, but rather a consequence of the \textit{processing methodology}. Our DWT implementation applied denoising \textit{globally} to the entire image. This global thresholding can inadvertently remove weak but important signal coefficients that are statistically similar to noise across the entire image, leading to a loss of fine detail and potential oversmoothing.

Conversely, the DFCT filter was applied using a \textit{block-based approach}, processing small ($16\times16$), overlapping partitions of the image independently. This method offers two key advantages:
\begin{enumerate}
    \item \textbf{Adaptation to Local Statistics:} Noise and signal characteristics can vary across an image. By processing small blocks, the DFCT filter effectively adapts to local image content and noise levels. A threshold that is optimal for a smooth region (high noise, low signal) may be different from one optimal for a textured or edge-rich region. Local processing allows for this adaptation, whereas a single global threshold in DWT cannot achieve the same flexibility.
    \item \textbf{Preservation of Local Details:} Operating on small blocks allows the DFCT to preserve fine details and textures within each block without being influenced by statistics from distant image regions. This prevents the blurring of edges and textures that can occur when a global operation averages out small-scale features.
\end{enumerate}
In essence, the block-based DFCT acts as an adaptive denoising method that respects the non-stationary nature of images. While the DWT's multi-resolution analysis is powerful, its effectiveness in this study was hampered by the use of a global processing strategy. The results suggest that for the additive noise models tested, the benefit of local adaptation in DFCT outweighs the benefit of the DWT's multi-scale edge representation when a global threshold is used.

This finding highlights a critical principle in image processing: the implementation strategy (global vs. local processing) can be as important as the choice of transform itself. Future work could explore \textit{local} or \textit{adaptive} wavelet thresholding schemes (e.g., context modeling, spatially varying thresholds) to leverage the strengths of wavelets while incorporating the local adaptability that made DFCT successful in this study. Figs. 8-11 show visual comparison of denoising results for the four kinds of noise. In the Figs. 8-11 the left image is noisy input image, the middle image is denoised image using the best wavelet method (BIOS wavelet with Hyperbola threshold for case of Gaussian and uniform noise, BIOS wavelet for Poisson noise and Daubechies wavelet with SURELevel for salt-and-pepper noise) and, the right image is denoised using DFCT.

\begin{figure}[htb]
\centering
\includegraphics[width=140mm,height=70mm]{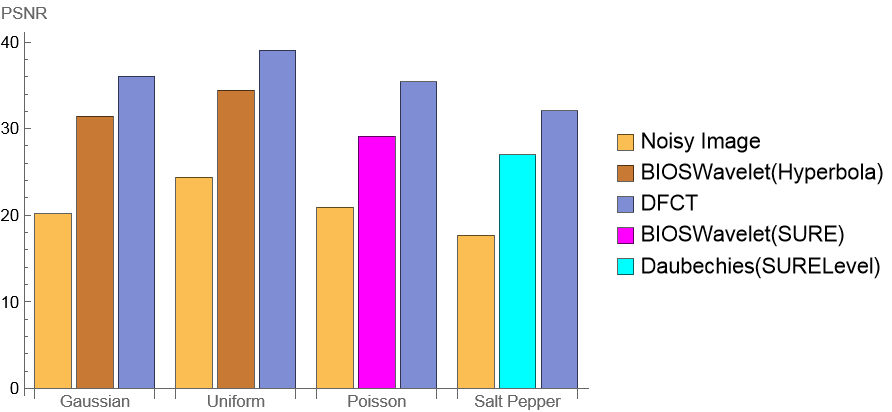}
\caption{Comparative performance analysis: Peak Signal-to-Noise Ratio (PSNR in dB) achieved by the best DWT method versus DFCT for all four noise types.} 
\end{figure}

\begin{figure}[htb]
\centering
\includegraphics[height=40mm]{gauss.png}
\includegraphics[height=40mm]{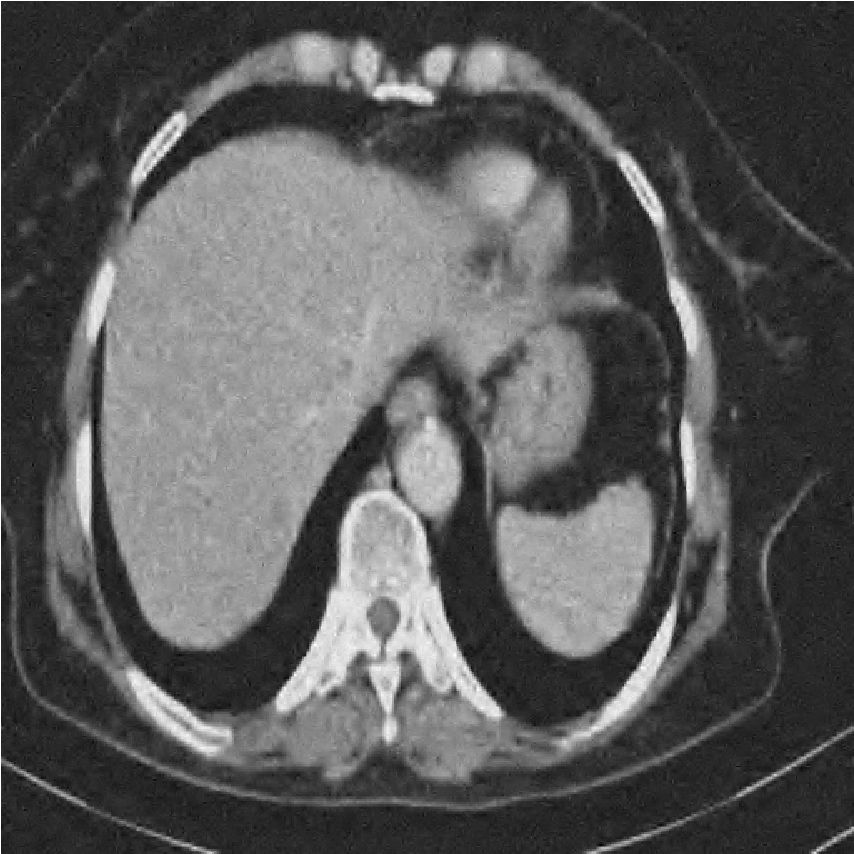}
\includegraphics[height=40mm]{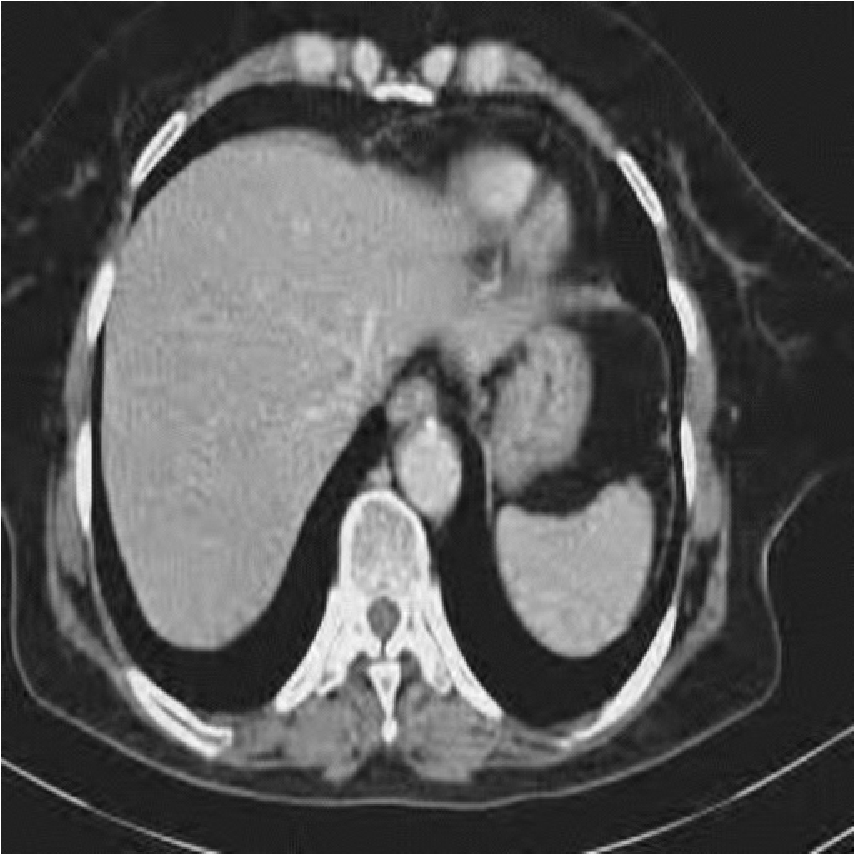}
\caption{Visual comparison of denoising results for Gaussian noise: (left) noisy input; (middle) denoised using the best wavelet method (Biorthogonal Spline wavelet with Hyperbola threshold); (right) denoised using DFCT.}		
\end{figure}
\begin{figure}[htb]
\centering
\includegraphics[height=40mm]{uniform.png}
\includegraphics[height=40mm]{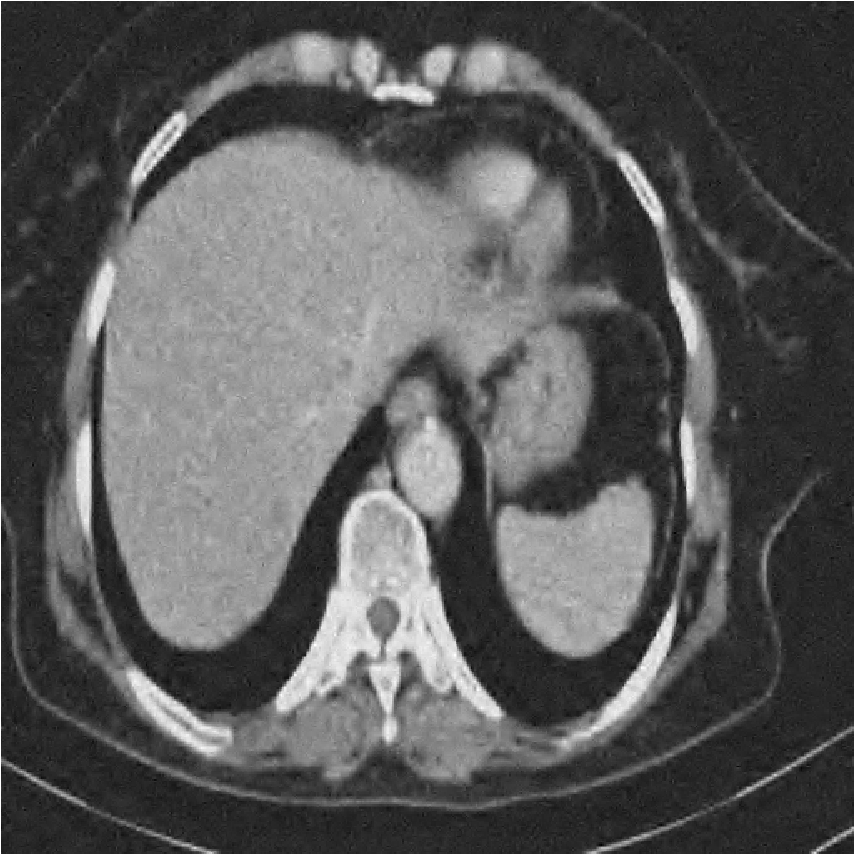}
\includegraphics[height=40mm]{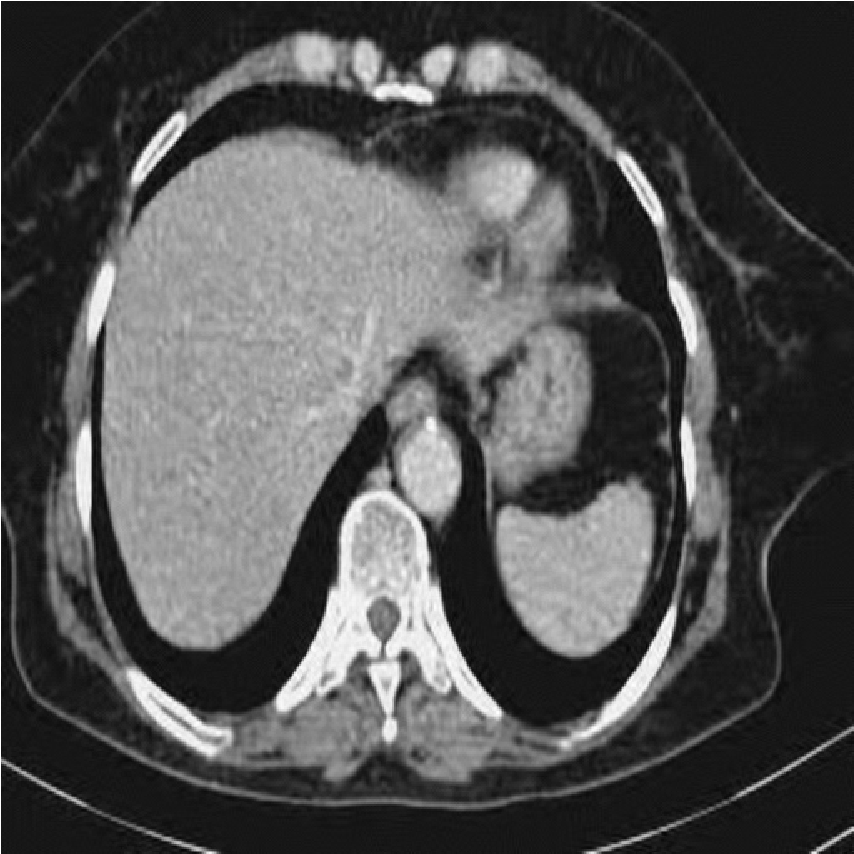}
\caption{Visual comparison of denoising results for uniform noise: (left) noisy input; (middle) denoised using the best wavelet method (Biorthogonal Spline wavelet with Hyperbola threshold); (right) denoised using DFCT.}		
\end{figure}
\begin{figure}[htb]
\centering
\includegraphics[height=40mm]{poisson.png}
\includegraphics[height=40mm]{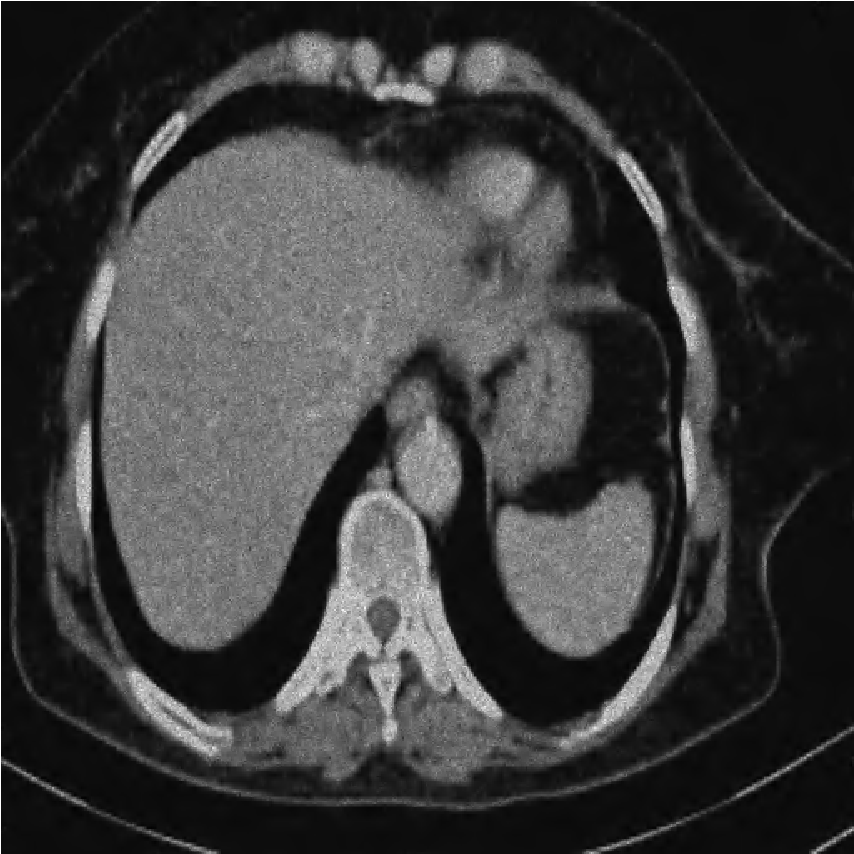}
\includegraphics[height=40mm]{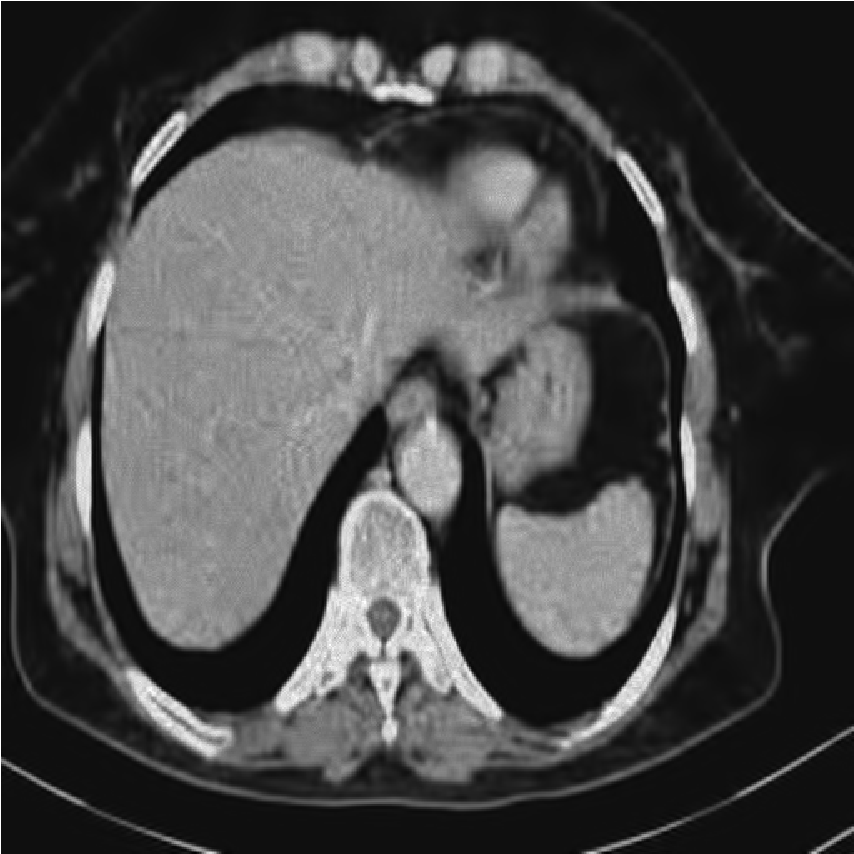}
\caption{Visual comparison of denoising results for Poisson noise: (left) noisy input; (middle) denoised using the best wavelet method (Biorthogonal Spline wavelet with SURE threshold); (right) denoised using DFCT.}		
\end{figure}
\begin{figure}[htb]
\centering
\includegraphics[height=40mm]{salt.png}
\includegraphics[height=40mm]{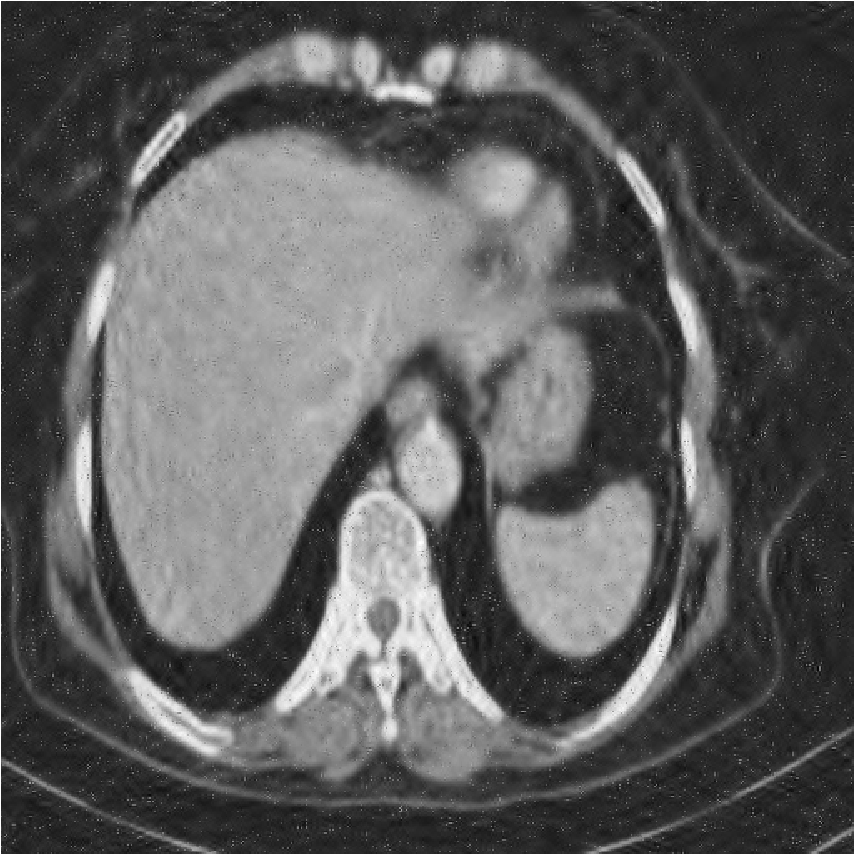}
\includegraphics[height=40mm]{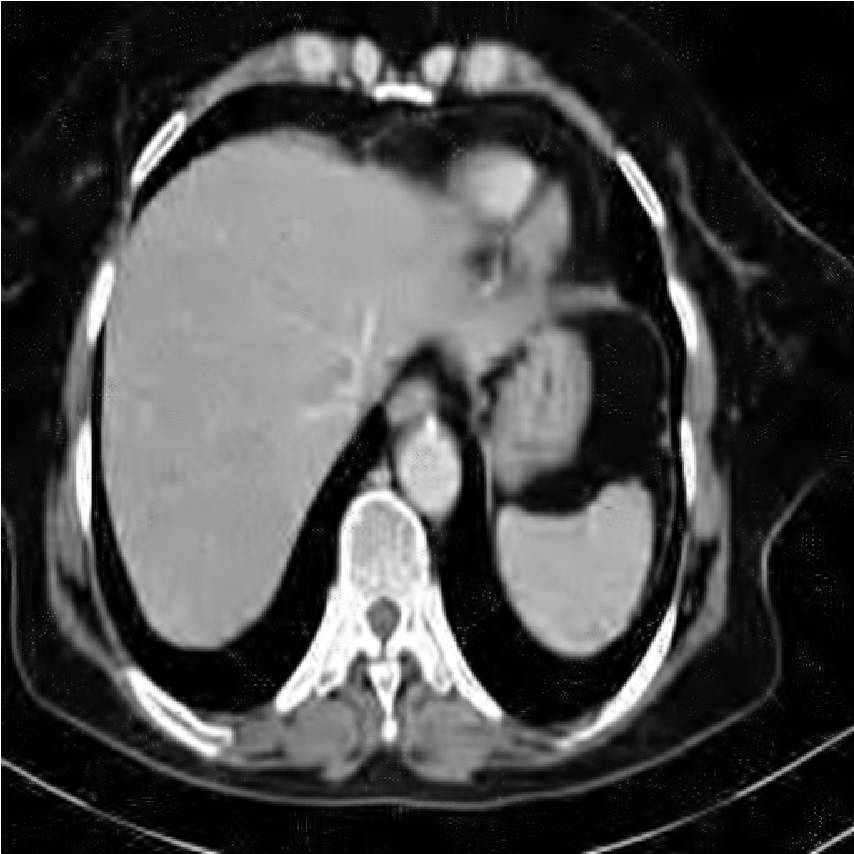}
\caption{Visual comparison of denoising results for salt-and-pepper noise: (left) noisy input; (middle) denoised using the best wavelet method (Daubechies wavelet with SURELevel threshold); (right) denoised using DFCT.}		
\end{figure}

\section{Conclusion}
This study conducted a thorough empirical comparison between Discret Wavelet Transform (DWT) and Discrete Fourier Cosine Transform (DFCT) filtering for denoising medical images corrupted with various types of additive noise. We evaluated a extensive suite of wavelet families (Haar, Daubechies, Coiflet, Symlet, CDF, Biorthogonal Spline, Meyer, Shannon) combined with numerous thresholding functions (Hard, Soft, Smooth Garrote, etc.) and selection rules (Universal, SURE, GCV).

The central and most significant finding is that a block-based DFCT filtering approach consistently and substantially outperformed a global DWT approach across all noise types (Gaussian, Uniform, Poisson, Salt-and-Pepper) and all quantitative metrics (SNR, PSNR, IM). The best-performing DWT methods achieved notable results, for instance, Biorthogonal Spline wavelets with hyperbola thresholding reached a PSNR of $34.46$ dB for uniform noise, yet were still surpassed by DFCT, which achieved a PSNR of $39.03$ dB for the same noise.

We conclude that the superior performance of DFCT is primarily attributable to its localized, block-based processing strategy. This approach allows the algorithm to adapt to the varying local statistics of an image, effectively separating noise from signal within small regions without the oversmoothing or loss of fine detail that can plague global denoising techniques like the DWT implementation used here.

Therefore, for practical medical image denoising applications where computational efficiency and effectiveness are paramount, block-based DFCT filtering presents a remarkably powerful and reliable solution. It offers a compelling balance of high performance, conceptual simplicity, and computational efficiency. This work underscores that the choice of denoising algorithm must consider not only the mathematical transform but also the processing framework (global vs. local). Future research should focus on developing and comparing adaptive, local-scale wavelet methods to fully leverage the potential of multi-resolution analysis in denoising tasks.

 \end{document}